\documentclass[aps,preprint]{revtex4}

\usepackage{amssymb}
\usepackage{amsmath}
\usepackage{graphicx}

\newtheorem{theorem}{Theorem}[section]
\newcommand{\be}{\begin{equation}}
\newcommand{\ee}{\end{equation}}
\newcommand{\ben}{\begin{eqnarray}}
\newcommand{\een}{\end{eqnarray}}

\begin{document} 
\title{Dimensional renormalizability in compactified spaces} 
\author{F.C. Khanna}
\affiliation{Theoretical Physics Institute, University of Alberta,
Edmonton, AB T6G 2J1, Canada\\
and TRIUMF, Vancouver, BC V6T 2A3, Canada}

\author{A.P.C. Malbouisson}
\affiliation{CBPF/MCT, Rua Dr. Xavier Sigaud, 150, Rio de Janeiro RJ, Brazil}

\author{J. M. C. Malbouisson}
\affiliation{Instituto de
F{\'\i}sica, Universidade Federal da Bahia, 40210-340, Salvador, BA, Brazil}

\author{A. E. Santana}
\address{{Instituto de F{\'\i}sica, Universidade de Bras{\'\i}lia, 70910-900,
Bras{\'\i}lia-DF, Brasil}}

\begin{abstract}

We first briefly review some aspects of the techniques of
dealing with ultraviolet divergences in Feynman amplitudes in an
Euclidian $D$-dimensional space-time. Next we consider
compactification of a $d$-dimensional ($d\leq D$) subspace. This
includes  effects of temperature and of  compactification of $d-1$
spatial coordinates. Then we show how dimensional renormalization
can be implemented for a field theory defined on this Euclidian
space-time with a compactified subspace.

\end{abstract}

\maketitle

\section{Introduction}

Studies on  field theories with compactified
 dimensions have their theoretical roots in the
finite temperature field theory historical procedure, of looking for
methods paralleling temperature-independent ($T=0$) theories, which
present practical and  well developed tools, as Feynman diagrams and
renormalization techniques. The first systematic approach to treat a
quantum field theory at finite temperature was presented in
1955~\cite{3mats1}, the Matsubara or \textit{imaginary-time}
formalism. Since then the development of the thermal field formalism
has followed the achievements of the $T=0$ quantum field theory. The
first generalization of the imaginary formalism was carried out
in in 1957~\cite {3ume4}, extending the Matsubara work to the
relativistic quantum field theory, and discovering  periodicity
(antiperiodicity) conditions for the Green functions of boson
(fermion) fields, a concept that later became known as the KMS
(Kubo, Martin and Schwinger) condition.

From a topological point of view, the Matsubara formalism is
equivalent to a path-integral evaluated on ${\mathbb R}^{D-1}\times
{\mathbb S}^{1}$, where ${\mathbb S}^{1}$ is a circumference of
length $\beta =1/T.$ As a consequence, the Matsubara prescription
can be thought, in a generalized way, as a mechanism to deal with
thermal effects and with spatial compactification. This concept has
been developed by considering a simply or non-simply connected
$D$-dimensional manifold with a topology of the type $\Gamma
_{D}^{d}={\mathbb R}^{D-d}\times {\mathbb S}^{1_{1}}\times {\mathbb
S}^{1_{2}}\cdots \times {\mathbb S}^{1_{d}}$, with ${\mathbb
S}^{1_{1}}$ corresponding to the compactification of the imaginary
time and $\,{\mathbb S}^{1_{2}},\dots, {\mathbb S}^{1_{d}}$
referring to the compactification of $d-1$ spatial
dimensions~\cite{NPB2002}. The topological structure of the
space-time does not modify the local field equations. However, the
topology implies  modifications of the boundary conditions over
fields and Green functions~\cite{ford1}. Physical manifestations of
this type of topology include, for instance, the vacuum-energy
fluctuations giving rise to the Casimir 
effect~\cite{casimir,mostep3,Jura1,Hebe,livro}, or in phase
transitions, the dependence of the critical temperature on the
parameters of compactification~\cite{livro,AMMS,linhares,JMP}.

In the topology $\Gamma _{D}^{d}$, the Feynman rules are modified by
introducing a generalized Matsubara prescription, performing the
following multiple replacements (compactification of a
$d$-dimensional subspace),
\begin{equation}
\int \frac{dk_1}{2\pi }\rightarrow \frac 1{\beta}\sum_{n_1=-\infty
}^{+\infty }\,,\;\;\;\;\int \frac{dk_i}{2\pi }\rightarrow \frac
1{L_i}\sum_{n_i=-\infty }^{+\infty }\;;\;\;\;k_1\rightarrow
\frac{2n_1\pi }{\beta}\;\;\;k_i\rightarrow \frac{2n_i\pi }{L_i}\;,
\label{Matsubara1}
\end{equation}
where $L_i\,,\,\,\;\;i=2,3...,d-1$ are the sizes of the compactified
spatial dimensions.

These ideas have had recently a regain of interest, particularly as
a new way to investigate the eletroweak transition and baryogenesis.
For instance a recent  investigation of the eletroweak phase
transition has been improved in \cite{pani,paniNPB} in the context of a
5-dimensional finite temperature theory with a compactified spatial
extra dimension. These authors conclude for a first-order transition
with a strength inversely proportional to the Higgs mass. Another
interesting result of~\cite{pani} is that up to temperatures of the
order of the inverse of the compactification lentgh, reliable (low
order) perturbative calculations lead to reasonable results. In
particular models where the Higgs field is identified with the
internal component of a gauge field in extra compactified dimensions
with size of inverse TeV~\cite{pani1} are considered. These are
known as models with gauge-Higgs unification, and are worked-out
examples~\cite{pani2,pani3,pani6,pani7}. Earlier references are
in~\cite{pani4} and an overview is found in~\cite{pani5}. The
five-dimensional (5D) case, with just one extra compactified
dimension, is the simplest one and also the one which seems
phenomenologically more appealing.

The situation summarized above leads to appropriate developments
in field theory on spaces with compactified dimensions, in
particular for implementing proper renormalization techniques in
such cases. We believe that a step in this direction is considered in this
paper, by setting a basis for  full development of renormalization
theory in space-time with spatial compactified dimensions, at zero
or finite temperature.

In the following, we first make a brief overview of the
fundamental aspects of renormalization theory in Sec.~\ref{sec2}, in order to 
make this article as self-contained as possible for a  
field-theorist reader. Then we show how
dimensional renormalization can be implemented in an Euclidian
space-time with a  compactified subspace.  For clear and rigorous
presentations of renormalization theory in non-compactified spaces, 
for both commutative and
non-commutative field theories, the reader is referred
to~\cite{rivasseau1,rivasseao,rivasseau2}.

\section{General aspects of perturbative renormalization} \label{sec2}

For definiteness we
consider the massive Euclidean $\lambda \phi^4_{D}$-theory described as usual, by 
the Lagrangian density,
\begin{equation}
\mathcal{L}=\frac{1}{2}\partial _{\mu}\phi (x)\partial ^{\mu}\phi
(x)+\frac{m^2}{2}\phi ^2(x) +\frac{\lambda}{4!}\phi ^4(x),
\label{Lag}
\end{equation}
in a non-compactified Euclidian $D$-dimensional space-time. 
In this case, 
the Feynman amplitude for a general diagrammatic insertion $G$ has an
expression of the form (omitting vertex factors and the overall
symmetry coefficient),
\begin{equation}
A_G(\{p\})\propto \int \prod
_{i=1}^{I_G}\frac{d^{D}q_{i}}{(2\pi)^D}\,\frac{1}{q_i^2+m^2}\prod
_{v=1}^{V_G}\delta \left(\sum_{j=1}^{I}\epsilon _{vj}q_j \right),
\label{AGgeraldelta}
\end{equation}
where $\{ p \}$ stands for the set of external momenta, $V_G$ is the
number of vertices, $I_G$ is the number of internal lines and $q_i$
is the momentum of each internal line $i$. The quantity
$\epsilon _{vi}$ is the {\it incidence matrix}, which equals $1$ if
the line $i$ arrives at the vertex $v$, $-1$ if it starts at $v$ and
$0$ otherwise. Performing integrations over the internal momenta using
the delta functions, it leads to a choice of independent loop-momenta
$\{k_{\alpha}\}$ and we get,
\begin{equation}
A_G(\{p\})= \int \prod _{\alpha
=1}^{L_G}\frac{d^{D}k_{\alpha}}{(2\pi)^D}\prod _{i
=1}^{I_G}\frac{1}{q_i^2(\{p\},\{k_{\alpha}\})+m^2}, \label{AGgeral}
\end{equation}
where  $L_G$ is the number of independent loops. The momentum $q_i$
is a \textit{ linear} function of the independent internal momenta
$k_{l} $ and of the external momenta $\{p\}$. By power counting, we
find that the integral in Eq.~(\ref{AGgeral}) is superficially
convergent if $DL_G-2I_G<0$; otherwise, if $DL_G-2I_G\geq 0$, the
integral is ultraviolet divergent. So, given a diagram $G$, we
define the quantity
\begin{equation}
d_G=DL_G - 2I_G \label{grau1}
\end{equation}
as the superficial degree of divergence of the diagram. If $d_G\geq
0$ the diagram will be ultraviolet divergent.

For any sub-diagram $S\subset G$ there are corresponding
sub-integrations, and we find that if
\begin{equation}
d_S=DL_S - 2I_S\geq 0, \label{grauS}
\end{equation}
where $L_S$ and $I_S$ are, respectively, the number of independent
loops and the number of internal lines of the sub-diagram $S$; an
ultraviolet divergence appears associated with the sub-diagram $S$.
Thus even if the diagram $G$ is superficially convergent, $d_G < 0$,
the Feynman integral $A_G$ can be divergent. For this, it is enough
that there is a sub-diagram $S$ such that $d_S\geq 0$. This has been
stated in Ref.~{\cite{zim1}}. A freely transposed version of this
statement is:

\begin{theorem}\label{cmasympt}
Let us consider a diagram $G$. If for all subdiagrams $S \subseteq G
$ we have $d_S<0$ the Feynman integral $A_G$ is ultraviolet
convergent. If there is at least one $S\subseteq G$, such that
$d_S\geq 0$, $A_G$ is ultraviolet divergent.
\end{theorem}
The divergent subdiagrams of a given diagram are called \textit{
renormalization parts}. For the full renormalization process, only
non-overlaping renormalization parts need to be
considered~\cite{bogren1,zim1}.

We present in the following an analysis in non-compactified spaces,
but the general features would apply as well in the case of a
compactified subspace, as it will be shown later. The basis  of the
perturbative renormalization method is that the starting theory is
not consistent as a physical model, and this fact manifests itself
as divergences. Then  attempts to modify the theory are made, by
introducing supplementary terms (\textit{counterterms}) in the
original Lagrangian, in such a way as to cancel the original
divergences.

An important step in dimensional renormalization  
is dimensional {\it regularization}. 
There are different regularization methods; all of them replace 
divergent Feynman amplitudes with more
general integrals by means of a set of supplementary parameters,
such that the theory does not have ultraviolet divergences when
these parameters belong to some domain. For a certain limit of these
parameters we find the original theory with their divergences. This
is a provisional procedure 
to explore more precisely the divergences to be suppressed in formal calculations. 
Some methods of regularization are:
 cutoff in the momenta,
Pauli-Villars regularization, analytic regularization, lattice
regularization and dimensional regularization~\cite{BG1,thoft12}. In
this case the idea is to define the Feynman integrals in a generic
space-time of dimension $D$ in such a way that the divergences are
recovered as poles of some functions. We will be particularly
concerned with the simple integral,
\begin{equation}
\int \frac{d^{D}q}{(2\pi )^{D}}\frac{1}{\left( q^{2}+M\right)
^{s}}=\frac{ \Gamma \left( s-\frac{D}{2}\right) }{(4\pi
)^{\frac{D}{2}}\Gamma (s)}\frac{1 }{M^{s-\frac{D}{2}}}\,.
\label{dimreg1}
\end{equation}
We will indicate symbolically a regularized amplitude as depending
on a parameter $\eta$ and the suppression of the regularization as
$\eta \rightarrow 0$. 
In what follows, unless explicitly stated, we understand 
{\it renormalized} quantities as the  limit $\eta \rightarrow 0$ of the 
properly regularized objects. 

For a renormalizable theory, we look for the organization of the set
of subtractions to be performed in order to define the complete set
of counterterms. When a counterterm $c_{S_{1}}$ for a subdiagram
$S_{1}$ with $ N_{1}$ external lines is present, the modified
Lagrangian contains a new vertex with $ N_{1}$ lines. For any
$S_{2}\supset S_{1}$, subtracting the divergent integration
corresponding to $S_{1}$ is equivalent to consider the sum
\[
A_{S_{2}}^{\rm{Ren}}=A_{S_{2}}+c_{S_{1}}A_{S_{2}/S_{1}},
\]%
where $A_{S_{2}/S_{1}}$ is the amplitude corresponding to the
reduced diagram obtained by shrinking the subdiagram $S_{1}$ to a
point. If $S_{2}$ is superficially divergent (independently of the
$S_{1}$-divergence), we must define another counterterm $c_{S_{2}},$
such that
\[
A_{S_{2}}^{\rm{Ren}}=A_{S_{2}}+c_{S_{1}}A_{S_{2}/S_{1}}+c_{S_{2}}.
\]
The process is continued in a recurrent manner, starting from the
smallest diagram to the larger ones. It may be shown that, in order
to obtain finite amplitudes, it is enough to take simultaneously all
the non-overlapping subdiagrams $ S.$ This is the origin of the BPH
(Bogoliubov, Parasiuk, Hepp) recursive
process~\cite{bogren1,zim1,hepp1}.

Having  defined all counterterms up to a given order $n$, the
renormalized amplitude for a diagram $G$ of the immediatly higher order,
$A_{G}^{\rm{Ren}},$ is given by,
\begin{equation}
A_{G}^{\rm{Ren}}=\sum _{\{S\}}\left[A_{G/\{S\}} \prod _{S\in
\{S\}}c_{S}\right]+c_{G}, \label{bogo}
\end{equation}
where $c_{G}$ is present if $G$ itself is  superficially divergent.
The sum in Eq.~(\ref{bogo}) is over all the families \{$S$\} of
superficially divergent non-overlapping subdiagrams of $G$,
including the empty family.
The amplitude $A_{G/\{S\}}$ corresponds to the diagram obtained by
reducing to a point each subdiagram of the family \{$ S$\}. In the
recursive process, it is understood that the intermediary step of
the regularization has been carried out, which is suppressed after the
recurrence is performed up to a given order. This procedure can be
generalized to take into account all renormalization parts of every
diagram $G$. Then we can state the {\it Bogoliubov-Parasiuk-Hepp
recurrence}~\cite{bogren1,hepp1} in the form,
\begin{theorem}\label{bogoth}
We define a {\it forest} ${\mathcal{F}}$ of $G$ as a set
$\{S_i\subseteq G\}$ of proper (connected and 1PI) subdiagrams such
that for $S_i,S_j\in \mathcal{F}$, either $S_i\subset S_j$,
$S_i\supset S_j$, or $S_i\cap S_j=\emptyset$. Then the renormalized
amplitude of the diagram $G$ can be written as,
\begin{equation}
A_{G}^{\rm{Ren}}=\sum _{\{S\}}\left[A_{G/\{S\}} \prod _{S\in
{\mathcal{F}}}c_{S}\right]+c_{G}, \label{bogo1}
\end{equation}
where $c_{G}$ is present if $G$ itself is  superficially divergent.
\end{theorem}

In Eq.~(\ref{bogo1}) the product of renormalization parts is to be
performed following the ordering in each forest, that is from
smaller to bigger diagrams. Therefore the renormalized amplitude may 
depend on the choice of momentum routing, that is, on the choice of
the independent loop momenta satisfying the $\delta$-function in
Eq.~(\ref{AGgeraldelta}). This difficulty leads to the definition
for each diagram, of  sets of {\it admissible} momenta and for
these, to the statement~\cite{zim1},

\begin{theorem}\label{bogoth1}
\it The amplitude $A^{\rm{Ren}}_G(\{p\})$ in Eq.~(\ref{bogo1}) is
convergent for any diagram $G$ in Euclidian space. Its
analytical continuation to the Minkowski space defines tempered
distributions.
\end{theorem}

An essential aspect of renormalization is to determine the {\it renormalization parts} 
of the theory under consideration, that is, how many
counterterms must be introduced in the theory to make it convergent.
For the $\lambda \phi_{D}^{4}$ model the superficial degree of
divergence is written as,
\begin{equation}
d_G=D-V_G(D-4)+N_G\left(1-\frac{D}{2}\right),  \label{d(G)1}
\end{equation}
where $N_G$ is the number of external legs. For $D=4$,  $d_G\geq 0,$ if, and only if, $N_G\leq 4.$ This implies
that to any order the only ultraviolet divergent diagrams will have
$ N_G=2,4$. From topological considerations, there are no diagrams
with $N_G=3$ in the $\lambda \phi ^{4}$ model.

The insertions $A_{G^{(2)}}$ and $A_{G^{(4)}}$ with 2 and 4 external
lines respectively, and only those, are ultraviolet divergent for
$D=4$. In this case we need to introduce only two counterterms in
the theory $c^{(2)}+c^{(2)\prime }$ and $c^{(4)}$ corresponding to
the diagrams with two and four external legs respectively.

The simplest case of {\it dimensional regularization} consists in
generalizing the formula given by Eq.~(\ref{dimreg1}) in dimension $D$ to a complex value $D^{\prime }$. This may be carried out for more
involved Feynman integrals, with the result that they become
meromorphic functions of $D^{\prime },\ A_{G}(D^{\prime })$, and the
ultraviolet divergences appear as poles of  Gamma- functions  at
$D^{\prime }=D$. The expansion around these poles allows us to define
the \emph{dimensional renormalization}: at each step in the
Bogoliubov-Parasiuk recurrence, we perform an expansion of the dimensionally regularized
amplitudes in powers of
$\epsilon =D^{\prime }-D$. {\it Dimensional renormalization} consists, essentially, in
subtracting the pole terms in the limit $\epsilon \rightarrow 0$, for each
renormalization part in the BPH recurrence.

The main advantage of dimensional renormalization is that, in
general, it respects the symmetry properties of the theory, which
are often dimensionally independent. On the contrary, in other
renormalization schemes, the
symmetry  usually needs to be re-established by adding new finite counterterms. 
In practical
applications dimensional renormalization must be carried out following the
BPH recurrence, step-by-step. An alternative procedure has been
found within the BPHZ (Bogoliubov-Prasiuk-Hepp-Zimmermann)
systematics~\cite{hepp1}, where an explicit global solution is
obtained for the dimensional renormalization~\cite{BD}. Other
rigorous renormalization procedures are given in
Refs.~\cite{BZ,BD,malbouisson,rivasseauren,malbouisson5}.

As far as the {\it definiteness of renormalization} is concerned, it is worth 
to recall that, whenever regularization is not suppressed, 
amplitudes  are finite to
a given perturbative order. Trouble starts when we suppress the
regulator. So, let us focus on regularized
objects, Feynman amplitudes, counterterms, etc..., emerging from
the bare Lagrangian density (\ref{Lag}). Two sets of counterterms, 
corresponding to two distinct renormalization schemes, 
 differ by a finite counterterm. To completely define the theory it is
essential to eliminate this ambiguity. This can be achieved by defining the
theory with physical conditions, fixing the normalization of
some Green functions at an arbitrary value of external momenta,
$\mu$. For the $\lambda \phi ^{4}_{4}$ theory it is
enough to fix the two- and four-point functions.
The renormalized Lagrangian density is obtained from the bare
Lagrangian by including counterterms,
\begin{equation}
\mathcal{L}^{\rm{Ren}}=\frac{\rm{Z}}{2}\partial _{\mu}\phi
\,\partial ^{\mu}\phi +\frac{\rm{Z}}{2}(m^2+c^{(2)})\phi ^2
+\frac{\rm{Z}^2 (\lambda +c^{(4)})}{4!}\phi ^4, \label{Lag1}
\end{equation}
where $\rm{Z}=\sqrt{1+c^{(2)\prime }}$. The 
counterterms $c^{(2)}$ and $c^{(4)}$ and $\rm{Z}$ and  are dependent on the regulator
$\eta $ and on the arbitrary parameter $\mu$. With the rescaling of
the field, $\bar{\phi }=\sqrt{\rm{Z}}\phi $ and defining the
physical mass and the renormalized coupling constant by $\bar{m}^2=m^2+c^{(2)}$
and $\bar{\lambda }=\lambda +c^{(4)}$ respectively, we have,
\begin{equation}
\mathcal{L}^{\rm{Ren}}=\frac{1}{2}\partial _{\mu}\bar{\phi
}\,\partial ^{\mu}\bar{\phi }+\frac{1}{2}\bar{m}^2\bar{\phi }^2
+\frac{\bar{\lambda }}{4!}\bar{\phi }^4. \label{Lag2}
\end{equation}
When the regularization is suppressed, everything diverges: 
counterterms and, for consistency, the bare mass and coupling
constant diverge, in such a way to provide
 $finite$
physical mass and coupling constant. The Lagrangian (\ref{Lag2})
generates perturbative series in the physical coupling constant
$\bar{\lambda}$. The independence of
physical quantities on the arbitrary mass parameter $\mu$ is
expressed by the well-known Callan-Symanzik equation~\cite{3pesq}.

\section{Compactification effects on renormalization}

\subsection{Compactification of imaginary time}

We now address the question about the renormalizability of a theory
at finite temperature. Specifically, we indicate how to use
dimensional regularization and analytic Zeta-function techniques to
calculate Feynman amplitudes at $T\neq 0$. Let us start with the
amplitude associated with a general diagram $G$ having $L$ internal
loops, given by Eq.~(\ref{AGgeral}). Using the identity
\begin{equation}
\frac{1}{Q_{1}\cdots Q_{I}} = \int_{0}^{1}dx_{1}\cdots
dx_{I}\,\delta \left(\sum\limits_{i=1}^{I}x_{i}-1\right)  \,
\frac{(I-1)!}{[x_{1}Q_{1}+\cdots+x_{I}Q_{I}]^{I}},
\label{identidade1}
\end{equation}
Eq.~(\ref{AGgeral}) can be cast in the form (from now on we suppress
the subscript $G$ from $L$ and $I$)
\begin{eqnarray}
A_{G}(\{p\}) & = & \int_{0}^{1}dx_{_{1}}\cdots
\int_{0}^{1}dx_{_{I-1}} \int \prod \limits_{\alpha =1}^{L}
\frac{d^{D}k_{\alpha }}{(2\pi)^{D}} \nonumber \\
 & & \times\, \frac{(I-1)!}{[x_{_{1}} q_{_{1}}^{2}+\cdots
+x_{_{I-1}} q_{_{I-1}}^{2} +(1 - \sum_{j=1}^{I-1}x_j)
q_{_{I}}^{2}+m^{2}]^{I}} \, ,   \label{V.5b}
\end{eqnarray}
where each $\,q_{i}\equiv q_{i}(\{p\},\{k_{\alpha}\})$ is a linear function of
the loop momenta $\{k_{\alpha}\}$. Now, completing squares, shifting
and then rescaling the integration variables, Eq.~(\ref{V.5b})
can be written in the form,
\begin{eqnarray}
A_{G}(\{p\})&=&\int_{0}^{1}dx_{_{1}}\cdots \int_{0}^{1}dx_{_{I-1}}\,
f_D(\{ x_j \})
\nonumber \\
 & & \times \int \prod \limits_{\alpha =1}^{L}
 \frac{d^{D}k_{\alpha }}{(2\pi
)^{D}} \, \frac{(I-1)!}{[k_{1}^{2}+\cdots
+k_{L}^{2}+\Delta^{2}]^{I}}, \label{AG2}
\end{eqnarray}
where $f_D(\{ x_j \}) = f_D(x_1,\dots,x_{I-1})$ and
\begin{equation}
\Delta^{2} = \Delta^{2}(\{p\},\{x_{j}\};m) = g(\{ x_j \})\, p^2 +
m^2 \label{Delta}
\end{equation}
is a function of the external momenta, $\{p\}$, of the Feynman
parameters, $\{x_{j}\}$, and of the mass $m$~\cite{3pesq}.

For an amplitude with $L$ independent loops, $A_{G}$, the Matsubara
prescription is applied to all $k_{\alpha}^{0}$ to get the
finite temperature expression,
\begin{eqnarray*}
A_{G}(\{p\};\beta )
&=&\frac{1}{\beta^L}\sum\limits_{\{l_{\alpha}=-\infty\}}^{\infty
}\int_{0}^{1}dx_{_{1}}\cdots \int_{0}^{1}dx_{_{I-1}} f_D(\{ x_j \})  \\
&&\times \int\prod\limits_{\alpha =1}^{L}\frac{d^{D-1}{\bf
k}_{\alpha }}{(2\pi )^{D-1}} \frac{(I-1)!}{[{\bf k}_{1}^{2}+\cdots
+{\bf k}_{L}^{2}+\sum_{\alpha=1}^{L}\frac{4\pi ^{2}
l_{\alpha}^2}{\beta ^{2}}+\Delta^{2}]^{I}}.
\end{eqnarray*}
We rewrite this equation as
\begin{equation}
A_{G}(\{p\};\beta) = \frac{1}{\beta^L}
\sum\limits_{\{l_{\alpha}=-\infty\}}^{\infty}
\int_{0}^{1}dx_{_{1}}\cdots \int_{0}^{1}dx_{_{I-1}} f_D(\{ x_j \})
B_{G}(\{p\},\{x_{j}\};\{ l_{\alpha} \},\beta), \label{AGB}
\end{equation}
where
\begin{equation}
B_{G}(\{p\},\{x_{j}\};\{ l_{\alpha} \},\beta) =  \int
\prod\limits_{\alpha =1}^{L}\frac{ d^{D-1}{\bf k}_{\alpha }}{(2\pi
)^{D-1}}\frac{(I-1)!}{ [{\bf k}_{1}^{2}+\cdots +{\bf
k}_{L}^{2}+\sum_{\alpha=1}^{L} b^2 l_{\alpha}^{2}+\Delta^{2}]^{I}},
\label{BGl}
\end{equation}
with
\[
b = \frac{2\pi}{\beta}\,.
\]

To perform the integration in Eq.~(\ref{BGl}), we proceed by
recurrence. We start by rewriting Eq.~(\ref{BGl}) as
\[
B_{G}(\{p\},\{x_{j}\};\{ l_{\alpha} \},\beta) = \int
\prod\limits_{\alpha =1}^{L}\frac{ d^{D-1}{\bf k}_{\alpha }}{(2\pi
)^{D-1}}\frac{(I-1)!}{[{\bf k}_{1}^{2}+\Delta _{1}^{2}]^{I}},
\]%
with $\Delta _{1}^{2}$ given by
\begin{eqnarray*}
\Delta _{1}^{2} & = & \Delta _{1}^{2}(\{p\},\{x_{j}\};
\{ l_{\alpha} \},m,\beta;\{ {\bf k}_{\alpha > 1} \})  \\
&=&{\bf k}_{2}^{2}+\cdots +{\bf k}_{L}^{2} + \sum_{\alpha=1}^{L} b^2
l_{\alpha}^{2} + \Delta^{2}(\{p\},\{x_{j}\};m).
\end{eqnarray*}%
Then, we  perform the integration over ${\bf k}_{1}$ by using the
formula given in Eq.~(\ref{dimreg1}) and obtain
\[
B_{G}(\{p\},\{x_{j}\};\{ l_{\alpha} \},\beta) = \frac{\Gamma
\left(I-\frac{D-1}{2} \right)}{(4\pi)^{\frac{D-1}{2}}} \int
\prod\limits_{\alpha =2}^{L}\frac{d^{D-1}{\bf k}_{\alpha }}{(2\pi
)^{D-1}}\, \frac{1}{[{\bf k}_{2}^{2} +
\Delta_{2}^{2}]^{I-\frac{D-1}{2}}},
\]%
where%
\begin{eqnarray*}
\Delta _{2}^{2} &=&\Delta _{2}^{2}(\{p\},\{x_{j}\};
\{ l_{\alpha} \},m,\beta;\{ {\bf k}_{\alpha > 2} \})\  \\
&=&{\bf k}_{3}^{2}+\cdots +{\bf k}_{L}^{2}+ \sum_{\alpha=1}^{L} b^2
l_{\alpha}^{2} + \Delta^{2}.
\end{eqnarray*}

The second step is to integrate over the momentum ${\bf k}_{2}$,
again using Eq.~(\ref{dimreg1}). The result is
\[
B_{G}(\{p\},\{x_{j}\};\{ l_{\alpha} \},\beta) = \frac{\Gamma
\left(I-2\left[\frac{D-1}{2}\right]
\right)}{(4\pi)^{2\left[\frac{D-1}{2}\right]}} \int
\prod\limits_{\alpha =3}^{L}\frac{d^{D-1}{\bf k}_{\alpha }}{(2\pi
)^{D-1}}\, \frac{1}{[{\bf k}_{3}^{2} +
\Delta_{3}^{2}]^{I-2\left[\frac{D-1}{2}\right]}},
\]
where
\begin{eqnarray*}
\Delta _{3}^{2} &=&\Delta _{3}^{2}(\{p\},\{x_{j}\};
\{ l_{\alpha} \},m,\beta;\{ {\bf k}_{\alpha > 3} \})  \\
&=&{\bf k}_{4}^{2}+\cdots +{\bf k}_{L}^{2}+ \sum_{\alpha=1}^{L} b^2
l_{\alpha}^{2} + \Delta^{2}(\{p\},\{x_{j}\};m).
\end{eqnarray*}
This procedure is continued until we have integrated over all
momenta. We end up with
\[
B_{G}(\{p\},\{x_{j}\};\{ l_{\alpha} \},\beta) = \frac{\Gamma
\left(I-L\left[\frac{D-1}{2}\right]
\right)}{(4\pi)^{L\left[\frac{D-1}{2}\right]}} \frac{1}{[\Delta
_{L}^{2}]^{I-L\left[\frac{D-1}{2}\right]} } \label{BGT}
\]
where
\begin{eqnarray*}
\Delta_{L}^{2} & = & \Delta_{L}^{2}(\{p\},\{x_{j}\};
\{l_{\alpha}\},m,\beta) \\
 & = & \sum_{\alpha=1}^{L} b^2 l_{\alpha}^{2} +
\Delta^{2}(\{p\},\{x_{j}\};m). \label{BGT1}
\end{eqnarray*}
The result for the amplitude then becomes
\begin{eqnarray}
A_{G}(\{p\};\beta ) & = & \frac{1}{\beta^L}\, \frac{\Gamma
\left(I-L\left[\frac{D-1}{2}\right]
\right)}{(4\pi)^{L\left[\frac{D-1}{2}\right]}}  \nonumber \\
&&\times \int_{0}^{1}dx_{_{1}}\cdots \int_{0}^{1}dx_{_{I-1}} f_D(\{
x_j \}) \sum\limits_{\{l_{\alpha}=-\infty\}}^{\infty}
\frac{1}{[\Delta _{L}^{2}]^{I-L\left[\frac{D-1}{2}\right]} }\, .
\label{AG1}
\end{eqnarray}

We recognize the sum over the set $\{l_{\alpha}\}$ in
Eq.~(\ref{AG1}) as one of the multi-variable Epstein-Hurwitz zeta
functions~\cite{Elizalde,Kirsten} defined by,
\begin{equation}
Z_{s}^{h^2}(\nu;a_1,\dots,a_s) =
\sum_{\{n_j=-\infty\}}^{+\infty}\frac{1}{\left( \sum_{r=1}^{s}
a_{r}^{2} n_r^2 + h^2 \right)^{\nu}}\, .
\end{equation}
This function can be analytically continued to the whole complex
$\nu$-plane, with the result~\cite{NPB2002},
\begin{equation}
Z_{s}^{h^{2}}(\nu ;\{ a_j \}) = \frac{\pi^{s/2}}{a_1 \cdots a_s\,
\Gamma(\nu)}\, \left[ \Gamma \left( \nu - \frac{s}{2} \right) h^{s -
2\nu} + F_{s} \left( \nu - \frac{s}{2};\{ a_j \},h \right) \right],
\label{Zs}
\end{equation}
where the function $F_{s} \left( \nu - s/2;\{ a_j \},h \right)$ is
the finite part, given by
\begin{eqnarray}
F_{s} \left(\nu - \frac{s}{2};\{ a_j \},h \right) &=& 4
\sum_{i=1}^{s} \sum_{n_i=1}^{\infty} \left( \frac{\pi n_i}{h{a_i}}
 \right)^{\nu-\frac{s}{2}} K_{\nu -\frac{s}{2}} \left( \frac{2 \pi h n_i}
 {{a_i}}
 \right) \nonumber \\
 & & +\, 8 \sum_{i<j=1}^{s} \sum_{n_i,n_j=1}^{\infty} \left(
 \frac{\pi}{h}\sqrt{\frac{n_{1}^{2}}{a_{1}^{2}} +
 \frac{n_{2}^{2}}{a_{2}^{2}}} \right)^{\nu-\frac{s}{2}} \nonumber \\
 & & \times K_{\nu -\frac{s}{2}}\left( 2 \pi
 h \sqrt{\frac{n_{1}^{2}}{a_{1}^{2}} + \frac{n_{2}^{2}}{a_{2}^{2}}} \right)
 \nonumber \\
 & &  +\, \cdots +\,  2^{s+1} \sum_{\{n_i\}=1}^{\infty} \left(
 \frac{\pi }{h}\sqrt{\frac{n_{1}^{2}}{a_{1}^{2}} +
\cdots + \frac{n_{s}^{2}}{a_{s}^{2}}}\right)^{\nu -\frac{s}{2}}
\nonumber \\
 & & \times \, K_{\nu -\frac{s}{2}}\left( 2\pi
h\sqrt{\frac{n_{1}^{2}}{a_{1}^{2}} + \cdots
+\frac{n_{s}^{2}}{a_{s}^{2}}}\right)  ,  \label{F}
\end{eqnarray}
and where $K_{\nu - s/2}$ denotes the modified Bessel function. The first
term in Eq.~(\ref{Zs}), proportional to $\Gamma \left( \nu - s/2
\right)$, has simple poles at $\nu = -n + s/2$, for $n\in
{\mathbb N}$.

Taking $s = L$, $\; a_1 = \cdots = a_L = b = 2\pi/\beta $, $\; h =
\Delta(\{p\},\{x_{j}\};m) \,$ and $\; \nu = I-L(D-1)/2$ in
Eqs.~(\ref{Zs}) and (\ref{F}), the $L$-loop amplitude,
Eq.~(\ref{AG1}), becomes
\begin{eqnarray}
A_{G}(\{p\};\beta ) & = & \frac{1}{2^{LD} \pi^{L(D-1)}} \left[
\Gamma\left( I - \frac{L D}{2} \right) \int_{0}^{1}dx_{_{1}}\cdots
\int_{0}^{1}dx_{_{I-1}}\, f_D(\{x_j\})\right. \nonumber \\
& & \left. \times \frac{1}{\left[
\Delta(\{p\},\{x_{j}\};m) \right]^{2 I - L D}}
\right. \nonumber \\
 & & + \left. \int_{0}^{1}dx_{_{1}}\cdots dx_{_{I-1}} \,f_D(\{x_j\})\,\right.
 \nonumber \\
& & \left. \times F_{L} \left( I-\frac{LD}{2};\{a_j=\frac{2\pi}{\beta}\}
,\Delta(\{p\},\{x_{j}\};m) \right) \right] . \nonumber \\
 & & \label{AG111}
\end{eqnarray}
The first term in this expression does not depend on the
temperature, $T=\beta^{-1}$, while the second term depends on the
temperature in such a way that it vanishes at zero temperature,
since $F_{L}\rightarrow 0$ as $T\rightarrow 0$ ($ \beta\rightarrow
\infty$). Furthermore, the first term (the $T=0$ contribution)
carries a singularity for space-time dimensions $D$ satisfying $I -
L D / 2 = 0,-1,-2,\dots$, while the temperature-dependent
contribution to the amplitude, the second term, is finite. To get
the renormalized amplitude, we have to suppress the singular part of
the first term and add its finite part to the second,
temperature-dependent, contribution. The singular part of the
amplitude is easily identified by expanding the $\Gamma$-function in
a Laurent series around the pole. The discussion presented so far is
restricted to the compactification of the imaginary time. It equally
applies to the compactification of one spatial coordinate in the
Euclidian $\lambda\phi^4$ theory. The generalization of this
procedure to the compactification of a subspace of dimension $d\subseteq D$
is presented in the following subsection.

\subsection{Finite temperature and spatial compactification}

The method of the previous subsection can be extended to the
case where, besides imaginary time, $d-1$ spatial dimensions
are also compactified. The set of compactification lengths will be
denoted by $\{ L_i \} = \{ L_1,\cdots ,L_{d-1} \}$, but no confusion
arises with the number of independent loops ($L$) of the diagram.
In this case, applying the generalized Matsubara prescription,
Eq.~(\ref{Matsubara1}), to Eq.~(\ref{AG2}) leads to
\begin{eqnarray*}
A_{G}(\{p\};\beta, \{ L_i \}) &=&\frac{1}{(\beta L_{1}\cdots
L_{d-1})^{L} }\sum\limits_{\{l_{(j)\alpha}=-\infty\}}^{\infty
}\int_{0}^{1}dx_{_{1}}\cdots \int_{0}^{1}dx_{_{I-1}} f_D(\{ x_j \})
\nonumber \\
& & \times \int\prod\limits_{\alpha =1}^{L}\frac{d^{D-d}{\bf
k}_{\alpha }}{(2\pi )^{D-d}} \frac{(I-1)!}{\left[{\bf
k}_{1}^{2}+...+{\bf k}_{L}^{2}+\sum_{\alpha=1}^{L} b_{j\alpha}^2
l_{(j)\alpha}^{2}+\Delta ^{2}\right]^{I}} \, ,
\end{eqnarray*}
where, now, ${\bf k}_{i}$ ($i=1,\dots,L$) are ($D-d$)-dimensional
vectors and we have numbered the Matsubara frequencies with integers
$l_{(j)\alpha}$ where $j=0,1,\dots,d-1$ refer to the compactified
coordinates and $\alpha$ has values from $1$ to $L$, the number of
independent loops of the diagram. Then the following steps are similar to
those leading to Eq.~(\ref{AG1}) and we get,
\begin{eqnarray}
A_{G}(\{p\};\beta ,\{ L_i \}) & = & \frac{1}{(\beta L_{1}\cdots
L_{d-1})^{L}  }\, \frac{\Gamma \left(I-L\left(\frac{D-d}{2}\right)
\right)}{(4\pi)^{L\left(\frac{D-d}{2}\right)}}  \nonumber \\
&&\times \int_{0}^{1}dx_{_{1}}\cdots \int_{0}^{1}dx_{_{I-1}} f_D(\{
x_j \}) \sum\limits_{\{l_{(j)\alpha}=-\infty\}}^{\infty}
\frac{1}{[\Delta _{Ld}^{2}]^{I-L\left(\frac{D-d}{2}\right)} },\nonumber \\
\label{AGd}
\end{eqnarray}
where
\begin{eqnarray*}
\Delta_{Ld}^{2}&=&\Delta_{Ld}^{2}(\{p\},\{x_{j}\};\{ l_{(j)\alpha} \},m,\beta) \\
 & = & \sum_{j=0}^{d-1} \sum_{\alpha=1}^{L} b_{j\alpha}^2 l_{(j)\alpha}^{2} +
\Delta^{2}(\{p\},\{x_{j}\};m) \label{Deltad}
\end{eqnarray*}
with $b_{0\alpha}=b=2\pi/\beta\,$ and
$\,b_{1\alpha}=2\pi/L_{1},\dots , b_{d-1,\alpha}=2\pi /L_{d-1}$ for
all $1 \leq \alpha\leq L$. The sum in Eq.~(\ref{AGd}) is the
multi-variable, ($d\times L$)-dimensional, Epstein-Hurwitz function,
\[
Z_{d L}^{\Delta ^{2}} \left( I-\frac{L(D-d)}{2};\{ b_{0\alpha} =
\frac{2\pi}{\beta} \},\{ b_{j\alpha} = \frac{2\pi}{L_j} \} \right)\,
,
\]
which possesses an analytical extension to complex values of $\nu =I-\frac{L(D-d)}{2}$ 
given by Eqs.~(\ref{Zs})
and (\ref{F}). Using these expressions, the regularized
finite-temperature amplitude, for $(d-1)$ ccompactified spatial
coordinates, is given by
\begin{eqnarray}
A_{G}(\{p\};\beta ,\{ L_i \}) & = & \frac{1}{2^{LD} \pi^{L(D-d)}}
\left[ \Gamma\left( I - \frac{L D}{2} \right)
\int_{0}^{1}dx_{_{1}}\cdots \int_{0}^{1}dx_{_{I-1}} f_D(\{ x_j
\})\right. \nonumber \\
& & \left. \times \frac{1}{ \left[ \Delta(\{p\},\{x_{j}\};m) \right]^{2 I - L D}}
\right. \nonumber \\
 & & + \int_{0}^{1}dx_{_{1}}\cdots \int_{0}^{1}dx_{_{I-1}} f_D(\{
x_j \})  \nonumber \\
 &&\left. \times\, F_{dL} \left( I-\frac{LD}{2}; \{ b_{0\alpha} \},\{ b_{j\alpha}
\},\Delta(\{p\},\{x_{j}\};m)
 \right) \right] .  \label{AGd11}
\end{eqnarray}

Again, the amplitude is separated into a zero-temperature free-space
contribution ($\beta , L_i \rightarrow \infty$), which eventually
has a singular part, and a contribution carrying the effects of
temperature and spatial compactification, which is finite. We then
state the theorem:

\begin{theorem}\label{dimrenS1}
Let us consider in the  $\phi ^{4}$ theory, a renormalization part,
a diagram $S\subseteq G$ belonging to a forest ${\mathcal{F}}$ of a
bigger diagram $G$, and its related finite-temperature amplitude,
with $(d-1)$ compactified spatial coordinates, $A_{S}(\{p\};\beta
,\{ L_i \})$. For the situations where $I - L D/2 = -n$,
$\,n=0,1,2,\dots $, the following quantity,
\begin{eqnarray}
A^{\rm{ren}}_{S}(\{p\};\beta ,\{ L_i \}) & = &
\frac{1}{2^{L D} \pi^{L (D-d)}} \nonumber \\
 & & \times \left[
\frac{(-1)^n}{n!}\psi(n+1) \int_{0}^{1}dx_{_{1}}\cdots
\int_{0}^{1}dx_{_{I-1}} f_D(\{ x_j \})\right. \nonumber \\
& & \left. \times \frac{1}{\left[
\Delta(\{p\},\{x_{j}\};m) \right]^{2 I - L D}}
\right. \nonumber \\
 & & + \int_{0}^{1}dx_{_{1}}\cdots \int_{0}^{1}dx_{_{I-1}}
 f_D(\{ x_j \})  \nonumber \\
 &&\left. \times\, F_{dL} \left(I-\frac{LD}{2};\{ b_{0\alpha}\},\{ b_{j\alpha}
\},\Delta(\{p\},\{x_{j}\};m)
 \right) \right],   \label{AGd11ren}
\end{eqnarray}
where the function $F_{dL}$ is the finite part of
\[
Z_{d L}^{\Delta ^{2}} \left( I-\frac{L(D-d)}{2};\{ b_{0\alpha} =
\frac{2\pi}{\beta} \},\{ b_{j\alpha} = \frac{2\pi}{L_j} \} \right)
\]
and $\psi(z)=d\ln\Gamma(z)/dz$, provides the {\it dimensionally
renormalized} amplitude of the diagram $S$, in what superficial
ultraviolet divergence in concerned. A similar statement
holds  for the reduced diagram $G/S$.
\end{theorem}

{\it Proof\,:} In Eq.~(\ref{AGd11}) divergences occur when
\[
I - \frac{L (D-d)}{2}=-n\,,\;\;\;n=0,1,2,\dots
\]
Then we use the Laurent expansion of the $\Gamma$-function around
its poles,
\begin{equation}
\Gamma (-n+\epsilon)= \frac{(-1)^n}{n!}\left[\frac{1}{\epsilon}+
\psi(n+1)+ {\mathcal{O}}(\epsilon)\right] \label{tad2}
\end{equation}
where $\psi(z)=d\ln\Gamma(z)/dz$, to subtract the poles of $\Gamma
\left(I-\frac{L(D-d)}{2}\right)$ in Eq.~(\ref{AGd11}). We are left
with the finite part $\frac{(-1)^n}{n!} \psi(n+1)$. This proves the
theorem. From Theorem~\ref{dimrenS1} the following theorem immediately follows. 

\begin{theorem}\label{bogoth2}
For all diagrams $G$ of the $\phi ^{4}$ theory,
Theorem~\ref{dimrenS1} ensures that $A_{G}^{\rm{Ren}}$ given by
Theorem~\ref{bogoth} is the {dimensionally renormalized} Feynman
amplitude of the diagram $G$ in a space-time with a compactified subspace.
\end{theorem}

This is easily proved since the result of theorem~\ref{dimrenS1} holds
for all renormalization parts $S$ of any diagram $G$. Then starting
from the smallest renormalization part $S$, which does not contain
any divergent subdiagram, the BPH recurrence in theorem~\ref{bogoth}
ensures the dimensional renormalization.
\\

\subsection{Examples}

We now proceed to present some examples. Consider first the one-loop
amplitude shown in Fig.~1, corresponding to the first correction to
the four-point function in the $\phi^4$ theory. This amplitude is
given by
\begin{eqnarray}
A_G & = & \int \frac{d^D k}{(2\pi)^D}\, \frac{1}{\left[ (p-k)^2 +
m^2
\right] (k^2 + m^2)} \nonumber \\
 & = & \int_{0}^{1} d x \int \frac{d^D k}{(2\pi)^D}\,
 \frac{1}{\left[ k^2 + x(1-x)p^2 + m^2 \right]^2} \, .
\end{eqnarray}
\begin{figure}[h]
\begin{center}
\includegraphics[{height=6.0cm,width=8.0cm}]{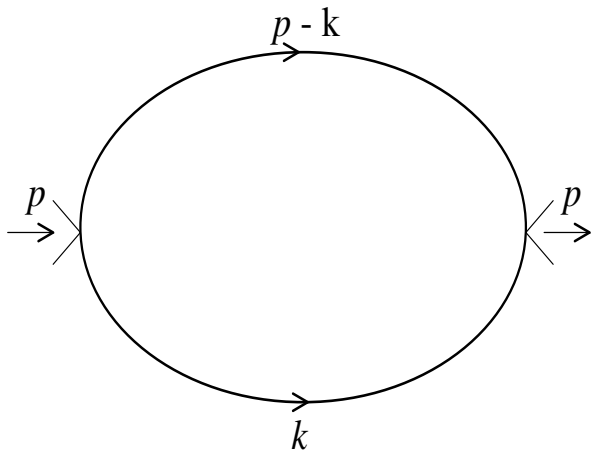} \caption{One-loop
contribution to the four-point function. } \label{figX2}
\end{center}
\end{figure}

\noindent {\it a}) {\it $D=4$, with one compactified spatial
dimension}:

\noindent For this one-loop case, $I - L D/2 = 0$ and the
subtraction of the pole term is required; we get
\begin{equation}
A^{\rm{Ren}}_{G}(p;L_1 ) = \frac{1}{16\pi^3} \left[ -\gamma + 4
\int_{0}^{1}\,dx
 \sum _{n=1}^{\infty} K_{0}\left( nL_1\sqrt{
x(1-x)p^{2} + m^2} \right)\right]\, , \label{D4}
\end{equation}
where we have used that $\psi(1)=-\gamma$, the Euler constant.

\noindent {\it b}) {\it $D=5$, with two compactified dimensions}
($\beta,L_1$):

\noindent Taking $D=5$ implies $I - L D/2 = -1/2$, and
Eq.~(\ref{AGd11}) gives directly a finite result,
\begin{eqnarray}
A^{\rm{Ren}}_{G}(p;\beta,L_1 ) & = & \frac{1}{32\pi^3} \int_{0}^{1}
dx \left[ -2\sqrt{\pi}\, \Delta(p;x,m) \right.
\nonumber \\
 & & +\, 4 \sum_{l=1}^{\infty} \left( \frac{2 \Delta}{\beta l}
 \right)^{\frac12} K_{\frac12}(l \beta \Delta) + 4 \sum_{n=1}^{\infty}
 \left( \frac{2 \Delta}{L_1 n} \right)^{\frac12} K_{\frac12}(n L_1
 \Delta) \nonumber \\
 & & \left. +\, 8 \sum_{l,n=1}^{\infty}
 \left( \frac{2 \Delta}{\sqrt{\beta^2 l^2 + L_1^2 n^2}}
 \right)^{\frac12} K_{\frac12}(\Delta \sqrt{\beta^2 l^2 + L_1^2
 n^2}) \right] \label{D5}
\end{eqnarray}
where
\[
\Delta(p;x,m) = \sqrt{x(1-2)p^2+m^2} \, .
\]
With either $\beta$ or $L_1$ going to infinity  the
amplitude reduces to that with only one compactified dimension.

For a two-loop example, consider the diagram of Fig.~2 which corresponds
to a second-order contribution to the propagator. In this case, we
 write
\begin{eqnarray}
A_G & = & \int \frac{d^D k}{(2\pi)^D}  \int \frac{d^D q}{(2\pi)^D}
\, \frac{1}{(k^2+m^2)[(q-k)^2+m^2][(p-q)^2+m^2]} \nonumber \\
 & = & \int_{0}^{1} dx \int_{0}^{1} dy \int \frac{d^D k}{(2\pi)^D}
  \int \frac{d^D q}{(2\pi)^D}
\, \frac{f_D(x,y)}{[k^2 + q^2 + g(x,y)p^2 + m^2]^3} \, ,
\end{eqnarray}
where
\begin{eqnarray}
f_D(x,y) & = & \frac{2}{\left[ (x+y)(1-y)-x^2 \right]^{D/2}}\; , \label{f5} \\
g(x,y) & = & \frac{xy(1-y)-yx^2}{(x+y)(1-y)-x^2}\; . \label{g2}
\end{eqnarray}
Taking $D=5$, in the present case, we obtain $I-LD/2=-2$, and so we
have to subtract the pole term of the $\Gamma$-function expansion.
Considering two compactified dimensions (the imaginary time, length
$\beta$, and a spatial coordinate, length $L_1$), the renormalized
amplitude is given by
\begin{eqnarray}
A^{\rm{Ren}}_{G}(p;\beta,L_1) & = & \frac{1}{2^{10}\pi^{6}} \left[
\frac{3-2\gamma}{4} \int_{0}^{1} dx \int_{0}^{1} dy f_5(x,y)
\Delta^4(p;x,y,m) \right. \nonumber \\
 & & +\, 32 \int_{0}^{1} dx \int_{0}^{1} dy f_5(x,y) \left\{
 \sum_{l=1}^{\infty} \frac{\Delta^2}{\beta^2 l^2} K_2(l \beta
 \Delta) \right. \nonumber \\
 & & + \sum_{n=1}^{\infty} \frac{\Delta^2}{L^2_1 n^2}
 K_2(n L_1 \Delta) + \sum_{l_1,l_2=1}^{\infty}
 \frac{\Delta^2}{\beta^2 (l^2_1+l^2_2)}
 K_2\left(\beta \Delta \sqrt{l^2_1+l^2_2}\right) \nonumber \\
 & & +\, 4 \sum_{l,n=1}^{\infty}
 \frac{\Delta^2}{\beta^2 l^2 + L^2_1 n^2}
 K_2\left(\Delta \sqrt{\beta^2 l^2 + L^2_1 n^2}\right) \nonumber \\
 & & + \sum_{n_1,n_2=1}^{\infty}
 \frac{\Delta^2}{L^2_1 (n^2_1+n^2_2)}
 K_2\left(L_1 \Delta \sqrt{n^2_1+n^2_2}\right) \nonumber \\
 & & +\, 4 \sum_{l_1,l_2,n=1}^{\infty}
 \frac{\Delta^2}{\beta^2 (l^2_1+l^2_2)+L_1^2 n^2}
 K_2\left(\Delta \sqrt{\beta^2 (l^2_1+l^2_2)+L_1^2 n^2}\right) \nonumber \\
 & & +\, 4 \sum_{l,n_1,n_2=1}^{\infty}
 \frac{\Delta^2}{\beta^2 l^2 +L^2_1 (n^2_1+n^2_2)}
 K_2\left(\Delta \sqrt{\beta^2 l^2 +L^2_1 (n^2_1+n^2_2)}\right) \nonumber \\
 & & +\, 4 \sum_{l_1,l_2,n_1,n_2=1}^{\infty}
 \frac{\Delta^2}{\beta^2 (l_1^2 + l_2^2) +L^2_1 (n^2_1+n^2_2)}  \nonumber \\
 & & \left. \left. \times \, K_2\left(\Delta \sqrt{\beta^2 (l_1^2 + l_2^2)
 + L^2_1 (n^2_1+n^2_2)}\right) \right\} \right]\, ,
\end{eqnarray}
where $f_5(x,y)$ is given by Eq.~(\ref{f5}) and $\Delta(p;x,y,m) =
\sqrt{g(x,y) p^2 + m^2}$, with $g(x,y)$ given by Eq.~(\ref{g2}).

\begin{figure}[h]
\begin{center}
\includegraphics[{height=6.0cm,width=8.0cm}]{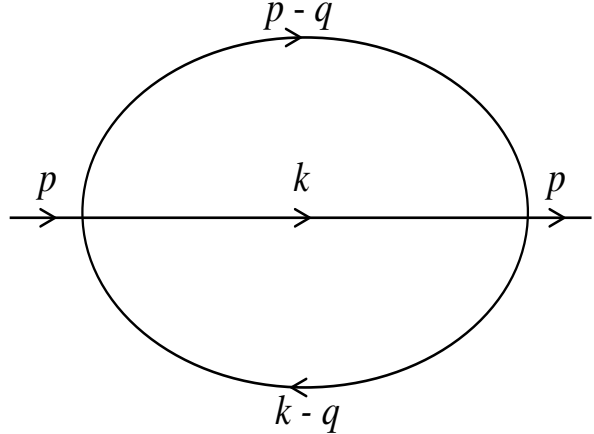} \caption{Two-loop
contribution to the propagator. } \label{figX3}
\end{center}
\end{figure}

\section{Concluding remarks}

The results in the previous section are obtained by the concurrent
use of dimensional and $zeta$-function analytic regularizations, to
evaluate  the integral over the continuous momenta and the summation
over the generalized Matsubara frequencies corresponding to the
compactified coordinates. Given a diagram $G$, ultraviolet
divergences arise from subdiagrams $S\subseteq G$ such that the
degree of divergence $d_S\geq 0$ in power-counting.  These
divergences appear as poles of $\Gamma$--functions with negative integer 
arguments (generally corresponding to even dimensions), a combination 
of the number of independent loops $L_S$, the number of internal
lines $I_S$ and of the space-time dimension $D$. 
Dimensional renormalization consists in extracting these poles,
which lead to counterterms to be inserted in the BPH
recurrence, given in theorem~\ref{bogoth}, to get finite,
renormalized quantities in a space-time with a compactified
subspace.

From a theoretical viewpoint, the general aspects of the topic 
presented here can be extended to
models where the matter field (bosons or fermions) is coupled with a
gauge field. In these theories, an important role is played by the
gauge symmetry in the discussion of perturbative renormalization.
The Ward-Takahashi relations, that manifestly contain the full
implications of the symmetry, have to be satisfied.

From a physical and phenomenological point of view, recently 
an interest in theories with  extra  compactified dimensions at the
inverse TeV scale arose in connection with the new LHC
(Large Hadron Collider) experiments. 
These theories provide a possible framework to throw some light on the
gauge hierarchy problem~\cite{pani}. Also, as we have mentioned before, a new idea
brought by theories with extra dimensions is the relation between 
the Higgs field and  the components of a gauge field. In the context of 
$5$-dimensions~\cite{pani,paniNPB}, the case of a scalar field coupled to a 
gauge field is considered, 
 where the non-vanishing component of the gauge field is  
along the compactified dimension. Models of this type are sometimes
called models with gauge-Higgs unification. Perhaps these theories provide an interesting framework
for physics beyond the standard model, even though numerous problems need to be solved.

{\bf Acknowledgments}

We are grateful to I. Roditi for discussions. This work  received
partial financial support from CNPq, FAPERJ (Brazil) and NSERC
(Canada).

\end{document}